\documentclass[aps,amsmath,amssymb,floats]{revtex4}
\usepackage{amsmath}
\usepackage{graphicx}
\usepackage{dcolumn}
\usepackage{bm}
\usepackage[dvips]{color}
\begin{document}

\title{The Rayleigh Quotient}
\author{M. K.G. Kruse\thanks{E-mail:kruse@physics.arizona.edu}$^a$, J. M.
Conroy\thanks{Justin.Conroy@fredonia.edu}$^b$ and H. G.
Miller\thanks{E-mail:hgmiller@localnet.com}$^b$}
\affiliation{$^a$Physics Department, University of Arizona, Tuscon, Arizona,
USA\\$^b$Physics Department, SUNY Fredonia, Fredonia, NY, USA}

\begin{abstract} 
The central role of the Rayleigh quotient in many body physics is
discussed. Various many body methods can be obtained from either an attempt to
evaluate the Rayleigh Quotient directly or through various variational
approximations. Rather than dwell on the
technical details necessary to obtain the equations of the various many body
methods, we  concentrate on how they can be obtained from the Rayleigh
Quotient, and some of the consequences of the approximations involved in their
evaluation.

\end{abstract}

\maketitle

Many of the approximation schemes employed in quantum mechanics can be gleaned
by simply considering the minimization of the Rayleigh quotient
\begin{equation}
 \delta(\frac{<\psi|\hat{H}|\psi>}{<\psi|\psi>})=0
\label{rq}
\end{equation}
where $\psi$ is the many body wave function and $\hat{H}$ is the many body
Hamiltonian operator. Clearly in order to perform the variation, the Rayleigh
quotient must exist and be finite.  This immediately rules out the possibility
of determining scattering states from equation (\ref{rq}) as they are not
normalizable in the usual $\mathcal{L}^2$ sense. Furthermore, if the operator,
$\hat{H}$, is not bounded in some manner  the Rayleigh quotient does not posses
a bound and the variational problem is no longer valid. 

In what follows, we shall assume that the Hamiltonian operator possesses at least
one bound, and for simplicity that it is bounded from below. If, for example,
$\hat{H}$ possesses only a point spectrum ({\em{i.e.}} only a bound state
spectrum) it must have at least a state of lowest energy (a ground state).
However, the space spanned by the many body wave functions need  not necessarily
be finite only denumerable. In this case, if no restriction is placed on the class of  many body trial functions other
than they be normalizable, equation(\ref{rq}) yields the many body Schroedinger
eigenvalue equation
\begin{equation}
 \hat{H}\psi=E\psi
\label{seq}
\end{equation} 
which is valid for every many body eigenfunction, $\psi$  whose corresponding
eigenvalue (or eigenenergy) is $E$.  The many body state with the lowest energy
corresponds to the ground state.

Note, however, that most many body Hamiltonian operators support not only a
bound state spectrum but also scattering states. The spectrum of bound states is
therefore not complete.
Although each eigenstate of such Hamiltonian operators may be expanded in a
complete set of basis states,
a simple unitary transformation between the eigenstates of said Hamiltonian
operator and any set of complete 
basis states does not exist. This, of course, may lead to problems if the 
solution of equation(\ref{rq}) (exact or approximate) is
expanded in an arbitrary set of basis states. Spurious solutions may
occur\cite{AM03,AM08,AMY09}.  

Now, in general, even if the space spanned by the many body eigenfunctions is
finite and denumerable, it is usually massive and a representation of the
operator $\hat{H}$ in terms of  many body basis states, yields a matrix which far
exceeds the storage capacity of even the largest computers. For this reason a
variety of approximation methods have been developed and often are referred to
as many body techniques. 

A conceptually simple way to make the problem tractable is to map the
Hamiltonian operator onto a small subspace in a manner first suggested by Bloch
and Horowitz\cite{BH58}.  The effective operator obtained should have the same
eigenvalues as the original Hamiltonian operator. Unfortunately the method is fraught with
difficulties.  The effective Hamiltonian is non-linear and no longer Hermitian 
and the resulting effective Schroedinger equation is no longer a simple
eigenvalue equation. Problems arising from truncation of the remaining part of
the many body space in the construction of this operator occur.  Furthermore the
solution of the effective Schroedinger equation yields only the projection of
many body eigenfunctions onto the small subspace, which necessitates the
construction of effective operators corresponding to any other
desired observables.  Improvements of the method have been made\cite{Sl80,S82}.

 A simple power  series expansion of the eigenpairs of $\hat{H}$ yields the
well-known perturbation expansion, which is a standard technique in quantum
mechanics, in spite of the fact that that the radius of convergence of this
series is often unknown. Even for the  simplest of potentials in one dimension (V$\propto \frac{1}{x^n}
\text{for n} \geq 5/2$) perturbative techniques fail, and more
complicated expansions involving re-summation techniques must be
used\cite{M94}. 

Rewriting 
equation (\ref{seq}) as an integral equation yields via Green's method,  iterative
schemes for obtaining the eigenpairs. Unlike as in the case of scattering, where
the energy in the Green's function is given, in the  bound state problems the
eigenenergy of the desired eigenstate must be known, in order to determine the
required Green's function.  
Simple techniques in 1D have been developed to  determine both the
eigenfunctions as well as the eigenenergies\cite{W98}. In the case of infinite
systems in 3D, more complicated techniques, which involve summations of the embedded two
particle
interactions, G-matrix expansions have been used with success. These have been
shown to be equivalent to coupled cluster expansions which have also been
employed in finite systems.

Techniques based on the Raleigh-Ritz method such as finite basis approximations
as well as the Lanczos algorithm\cite{L50} have also been employed.  In this
case a small number of basis states are selected and the Hamiltonian operator is
diagonalized in this subspace.  In calculations performed in atomic physics sum
rule considerations are used as a criterium to select the space spanned by
these basis states\cite{GD82}.  In the case of the Lanczos algorithm\cite{D98} an
orthonormalized set of Krylov vectors\cite{P80} 
is used to construct iteratively a matrix represention of $\hat{H}$ which
is then diagonalized. However, as has been pointed out, spurious solutions may
occur\cite{AM08}, but can be identified easily\cite{AM03,AMY09}.

Alternatively, approximations to the many body wave function may be made either
before, or after the variation of the Rayleigh quotient.  Restricting the class
of trial wave functions before variation leads to many of the standard
variational
techniques developed in many body physics.  For example, restricting the trial
many body wave function to a single Slater determinant (an antisymmetrized
product of single particle orbitals or wave functions) yields the well known
Hartree Fock (HF) equations. In spite of the fact that the variation is a linear
operation, the fact that it it is applied in a product space of single particle
orbitals, leads to  a non-linear set of equations, which must be solved
iteratively. Although the solution of the HF equations corresponding to the
ground state is often obtained by diagonalizing the non-linear HF Hamiltonian
operator, no mathematical proof of convergence exists. However, considerable
numerical evidence exists that this algorithm is convergent and yields a simple
approximation of the exact ground state.  More complicated choices of the
trial wave function lead to the multi-configurational HF equations and the
introduction of quasi-particles into the ansatz for the orbitals used in the
trial wave function lead to other approximation schemes, for example the   HFB
and the BCS approximation schemes.

An additional approximation may be made by using a finite basis expansion for
the variational
state. In the HF case, one obtains the Hartree-Fock-Roothaan (HFR)
equations\cite{R51}. While it is
clear that Hartree- Fock single particle orbitals  can be expanded in an
appropriately chosen finite set of basis states,
an {\em a priori} choice of basis states need not necessarily lead to the same
ground state solution of the HF and HFR equations.

In both finite basis approximations, as well as in variational methods such as
HFR, the choice of the set of basis states is extremely important. In atomic
physics, where finite basis approximations are  employed,
sum rule considerations are routinely used to select the space spanned by the
basis functions\cite{GD82}. In nuclear physics,
the presence of low-lying collective states, strongly suggests that relatively
few properly chosen basis states are required\cite{MS71,Dytrych07}.
This being the case, it has been suggested that one could use extended variational
methods, based on the HF approximation, to generate a set of orthogonal basis
states to use in a Rayleigh Ritz calculation of the exact
eigenstates\cite{MDD78,MD79}.

On the other hand, one might try to simply evaluate the Rayleigh quotient exactly
using path integrals. The path integral approach to many body systems typically
involves the selection of an overcomplete set of
states\cite{K60,K80,B81,B80,KLA1}. 
For example, coherent states are a particularly useful choice, since they are
eigenstates of the annihilation operator, thus simplifying the evaluation of
matrix elements.  Coherent states also yield a resolution of unity that makes
constructing many body functional integrals simpler.  Recent progress has been
made regarding the efficiency of numerical evaluations of these path integrals 
(e.g. \cite{BV1,BV2}). For a particular
overcomplete set of states, classical equations of motion can be obtained using
the standard saddle-point approximation.  Different choices in the overcomplete
set produce different classical approximations.  For example, by choosing the
set of all Slater determinants, one obtains the Time-dependent Hartree-Fock
Equations.  As pointed out in \cite{B81}, the choice of a particular
overcomplete set is equivalent to a particular choice of trial wavefunctions in
the variational principle. Moreover, the functional integral approach provides a
way of calculating corrections to the classical approximation.

Alternatively one might use Monte Carlo techniques to evaluate the multi
dimensional integrals in the Rayleigh quotient. In the earlier Shell Model Monte
Carlo (SMMC)method\cite{KDL97} statistical mechanical techniques were used to
to reduce the imaginary many body evolution operator to a coherent
superposition of fluctuating one body fields. The resulting path integral is
evaluated stochastically. Difficulties arise when the weights
used in the evaluation of observables are not positive. This sign problem
occurs for many body Hamiltonians used in nuclear physics calculations.
In the Quantum Monte Carlo Diagonalization\cite{OMH99}, stochastical sampling of
the many body states is used to select only those states, which are important to
the eigenstate to  be obtained.  The ground state is obtained
by performing a diagonalization of the Hamiltonian operator in this basis.

Full configuration interaction methods (full CI), in which a single reference
state is used, from which all $n$-tuple excitations are generated, are always
one of
the best possible many-body calculation that one can do, but unfortunately is
limited by current computational abilities. However, there are techniques which
approximate full CI calculations, one of which is the method of coupled cluster
\cite{CCrev07}. Coupled cluster methods also have the advantage that the
calculated energies scale properly with system size, ({\em i.e.}, linearly in
particle number). This property is known as size-extensivity. In most many body
calculations, the idea of size-extensivity, as well as
the closely related property of size consistency is often overlooked. These two ideas
are related, and so, to avoid confusion, we will briefly describe them.
Size-consistency is most easily
understood from the viewpoint of a chemical reaction. The dissociation of a
single system, ``AB'', into two parts, system A and B, which eventually
are infinitely separated from each other, should give the same total energy as
the original system ``AB''. In other words, $E_{AB}=E_A+E_B$. To mathematically
define size consistency is very difficult, and arguments as to whether quantum
systems are ever truly separated arise \cite{Duch94}. Size extensivity on the
other hand,
states that the energy of a system, such as an electron gas, should scale
linearly with the number of particles present. Such an idea is much easier to
formulate mathematically. In terms of approximation methods used in many-body
 techniques, size extensivity is discussed much more frequently, and is often
the most desired property, whereas size consistency is usually an afterthought.
The most logical step is to limit CI calculation to single and double
excitations, often abbreviated as CI-SD. In
this case, the CI-SD wavefunctions are generated as follows, $\Psi_{\rm
CISD}=(1+\hat{C_1}+\hat{C_2})\Phi_0$, in which $\Phi_0$ is the single Slater
determinant used as a reference function, and the $\hat{C_i}$ are the operators
that generate the single and double excitation on top of the reference function.

Unfortunately this truncation scheme lacks size extensivity. This effect was
first noticed by Brueckner in studies of the electron gas \cite{Brueckner55},
where the
number of particles $N$ goes to infinity. The energy should be linear in $N$,
but terms arising from the Raleigh-Schroedinger perturbation theory (RSPT)
expansion contained terms that were proportional to $N^2$ and $N^3$. Brueckner
showed that these unphysical terms are canceled up to fourth order. Goldstone
demonstrated that these unphysical terms cancel to all orders
\cite{Goldstone57}, since
the perturbation terms can be decomposed into linked and unlinked diagrams. The
unlinked diagrams are the terms that destroy size extensivity, but provided one
does RSPT to infinite order, are always canceled out. This is the linked
diagram theorem. CI calculations are related to RSPT expansions, in order to
extract the CI eigenvalues. In the case of CI-SD, the calculation retains these
unlinked diagrams that are proportional to $N^2$ etc, and are thus not canceled
out. The cancellation occurs if one were to add
more excitations to the CI calculation, such as CI with singles, doubles,
triples and quadruples included. Unfortunately, unlinked diagrams will still
remain, since they are canceled by an ever higher order of excitation. The
complete cancellation of unlinked diagrams only occurs once all excitations are
included, but this brings us back to the computational problems of full CI.

Size extensivity can be restored in the RSPT expansion. The extensivity is
restored if one considers all configurations to a given order. This leads to
many-body perturbation theory (MBPT), which is a fully-linked diagrammatic
expansion, order by order, and thus size extensive, up to that given order
\cite{Bartlett78,CCrev07}.
Unfortunately, each subsequent order is more difficult to calculate than the
previous one, rendering the method useful, but not efficient. Coupled cluster
(CC) offers a slightly different approach to the problem, by providing an
infinite order resummation of MBPT in selected clusters, such as single and
double excitations. CC is by construction size extensive. Although CI and CC
both have the same inherent ideas, {\em i.e} generate single and double
excitations on top of a reference state, the difference to CI comes from the
exponential ansatz made in CC. The coupled-cluster wavefunction is generated by
$\Psi_{\rm CC}=exp(T)\Psi_0$, where $T=T_1+T_2+\ldots+T_n$, where $T_p$ is a
connected cluster operator that generates the $p$-fold excitation. It is this
exponential form which ensures the size extensivity of the method. In the case
of single and double excitations, one speaks of CCSD calculations. CCSD actually
builds in more correlations than CISD would, since the exponential expansion
generates terms such as $T^2/2$ and $T_1T_2$. Although CC has some advantages
over CI, such as being size extensive and more efficient, it is not as
versatile as truncated CI. In the case of nuclear structure, CC is often used to
calculate the ground state energies of doubly-magic nuclei, such as $^{40}$Ca
\cite{Hagen07}. Recently, CC in nuclear structure has been extended to $A\pm2$
nuclei, in which $A$ represents a doubly magic nucleus \cite{Jansen11}. For
other nuclei, the
techniques of CI are commonly used. 

In the preceding few paragraphs, we have discussed the role of size extensivity
in many-body calculations. Although CC is size extensive, it does suffer from
another difficulty, related to the truncation of the basis and the introduction
of contamination of the many-body wavefunction with center of mass motion. Both
CI and CC have their basis truncated at the single particle level. On the other
hand, the basis used in the No-Core Shell Model (NCSM) \cite{Nav09}, a method
for calculating
observables of light nuclei from an {\em ab-initio} viewpoint, has its basis truncated
at the level of energy quanta. In nuclear structure, CI, CC and the NCSM all use
as single particle states the harmonic oscillator functions, $\psi_{nljm}(r)$.
The single particle states are antisymmetrized to form Slater determinants,
forming the many-body basis, in which the diagonalization is done. Returning to
the two different truncation
schemes, in CI-SD and CCSD, two particles could be placed in single-particle
states that correspond to the highest single-particle energy possible. For
example, if we truncate our basis at
the $2\hbar\omega$ level on top of a reference state build up of only $0s$
components, we could place our two particles in any of the $sd$-shell single
particle states. On the other hand, in the NCSM, the truncation is defined by
sharing $2\hbar\omega$ units of energy between the $A$ nucleons present. In this
case, we could place our two particles in the $p$-shell, or place one particle
in the $sd$-shell, and the other particle in the $s$-shell. This subtle
difference has implications for the separation of center of mass and intrinsic
states. 

The nuclear Hamiltonians that are solved in nuclear structure are initially
translationally invariant. However, due to our methods of solving the
Hamiltonian in a single particle basis (and not relative coordinates), we
introduce states that correspond to the center of mass motion of the nucleus.
These states can be separated out by the Lawson projection, if the
many-body wavefunction factorizes into an intrinsic and center of mass term,
$\Psi(\vec{x},\vec{X_{cm}})=\psi(\vec{x})\otimes\psi(\vec{X_{CM}})$, in which
$x$ represents the $3(A-1)$ position coordinates and $\vec{X_{CM}}$ represents
the
center of mass coordinate. This factorization is guaranteed when the basis is
truncated as it is in the NCSM, but not when the truncation is performed at the
single particle level. The single particle harmonic oscillator functions are
related to relative and center of mass coordinates through the
Talmi-Brody-Moshinsky brackets \cite{TM60}, which is an orthogonal transformation between
the two  coordinate systems. However, the transformation involves the sum over
expansion
coefficients, in which the sum is restricted to the maximal energy quanta two
nucleons may share among each other. In the case of CC, the question of center
of mass contamination has been recently investigated, and a factorization has
been found, although it is not clearly understood \cite{Hagen09}. 

When we relate this to the Raleigh quotient, we find that although approximate
methods can be used, even with proper extensive scaling, their mere formulation
on the number of excitations built into the technique (such as singles and
doubles etc), is not sufficient to converge to the true eigenvalues of the
Hamiltonian; one needs to consider the effects of the actual basis too, since
substantial error can occur, such as those caused by center of mass
contamination.

In this brief note we have attempted to  point out the central role of the
Rayleigh Quotient in many body physics.  Rather than dwelling on the 
technical details necessary to obtain the relevant  equations of the various
many body
methods, we have concentrated on how they can be obtained from the Rayleigh
Quotient, and some of the consequences of the approximations involved in their
evaluation.

\end{document}